\documentclass[aps,pre,twocolumn,amsmath,showpacs,superscriptaddress]{revtex4}

\usepackage{amsmath}
\usepackage{amsbsy}
\usepackage{amscd}
\usepackage{amsopn}
\usepackage{amstext}
\usepackage{amsxtra}
\usepackage{epsfig}
\usepackage{enumerate}
\usepackage{hyperref}

\def\al{\alpha}
\def\bt{\beta}
\def\gm{\gamma}
\def\dl{\delta}
\def\ep{\epsilon}
\def\sg{\sigma}
\def\tu{\tau}
\def\la{\langle}
\def\ra{\rangle}
\def\nn{\nonumber}

\def\wt{\widetilde}

\begin{document}
\title{Cascades on a class of clustered random networks}
\author{Adam Hackett}
\affiliation{MACSI, Department of Mathematics \& Statistics, University of Limerick, Ireland.}
\author{Sergey Melnik}
\affiliation{MACSI, Department of Mathematics \& Statistics, University of Limerick, Ireland.}
\author{James P. Gleeson}
\affiliation{MACSI, Department of Mathematics \& Statistics, University of Limerick, Ireland.}
\date{\today}


\begin{abstract}
We present an analytical approach to determining the expected cascade size in a broad range of dynamical models on the class of random networks with arbitrary degree distribution and nonzero clustering introduced in [M.E.J. Newman, Phys. Rev. Lett. 103, 058701 (2009)]. A condition for the existence of global cascades is derived as well as a general criterion which determines whether increasing the level of clustering will increase, or decrease, the expected cascade size. Applications, examples of which are provided, include site percolation, bond percolation, and Watts' threshold model; in all cases analytical results give excellent agreement with numerical simulations.
\end{abstract}

\maketitle

\section{Introduction}\label{sec1}
The network topologies on which many natural and synthetic systems are built provide ideal settings for the emergence of complex phenomena; see the reviews \cite{Newman03,DorogovtsevMendes03,AmaralOttino04,BoccalettiEtAl06,Newman10}. One well-studied manifestation of this, called a \emph{cascade} or \emph{avalanche}, is observed when under certain circumstances interactions between the components of a system allow an initially localized effect to propagate globally. For example, the malfunction of technological systems like email networks or electrical power grids is often attributable to a cascade of failures triggered by some isolated event. Similarly, the transmission of infectious diseases and the adoption of innovations or cultural fads may induce cascades amongst people in society.

It has been extensively demonstrated \cite{Watts02,HolmeKim02,MotterLai02,MorenoEtAl03,GohEtAl03,DoddsWatts04,CrucittiEtAl04,LaiEtAl05,LeeEtAl05,GohEtAl05,DuanEtAl05,AntalEtAl06} that the dynamics of cascades depend sensitively on the patterns of interaction laid out by the underlying network. One of the goals of network theory is to provide a solid theoretical basis for this dependence. In order to do this one must first construct network models which are both mathematically sound and which capture the salient features of their real-world counterparts. So far there has been limited success in this direction. Most existing analytical results derive from the class of random networks defined by the so-called \emph{configuration model} \cite{BenderCanfield78,Bollobas80}. The degree distribution of a network, $p_{k}$, specifies the fraction of its nodes (vertices) that have $k$ incident edges. In the configuration model, one generates a network of size $n$ and given $p_{k}$ by attaching, with appropriate probabilities, $k$ \emph{stubs} to each of a set of $n$ nodes, and then randomly connecting pairs of these stubs together to make complete edges. The major shortcoming of this approach is that in the limit $n\rightarrow\infty$ the density of cycles of length three (triangles) in the resulting network will vanish. In contrast, it is well established that the presence of closed interactions in real-world networks engenders significant numbers of these short cycles. This feature is usually quantified using some version of the \emph{clustering coefficient}, which has been described in a sociological context as the probability that ``the friend of my friend is also my friend'' \cite{WattsStrogatz98}.

Recently, Newman \cite{Newman09} and Miller \cite{Miller09a} independently proposed an extension to the classical configuration model to include nonzero levels of clustering (even as $n\rightarrow\infty$), thus opening the doors to the derivation of new analytical results for cascade dynamics on somewhat more realistic network topologies. Newman's model (which is the primary focus of our investigation) introduces a joint distribution, $p_{st}$, specifying the fraction of nodes that are each connected to $s$ single edges and $t$ triangles, thereby directly embedding triangles of interconnected nodes into a locally tree-like structure. Since the parameter $t$ controls the density of triangles it also determines the clustering coefficient. In addition, it was shown in \cite{Newman09} how the generating function formalism of \cite{NewmanEtAl01} can be applied to these networks to derive expressions for some of their fundamental properties.

In this paper we demonstrate an analytical approach to determining the mean cascade size in a broad range of dynamical models on the clustered random networks of \cite{Newman09}. This approach extends the work of Gleeson and Cahalane \cite{GleesonCahalane07} and Gleeson \cite{Gleeson08} on locally tree-like networks which itself was built on methods introduced to study the zero-temperature random-field Ising model on Bethe lattices \cite{SethnaEtAl93,DharEtAl97,Shukla03}. We consider a specific class of models which satisfy the following  properties: (i) each node is assigned a binary value specifying its current state, \emph{active} (\emph{damaged} or \emph{infected}) or \emph{inactive} (\emph{undamaged} or \emph{susceptible}); (ii) the probability of a node becoming active (in a synchronous update of all nodes) depends only on its degree $k=s+2t$ and the number $m$ of its neighbors who are already active, this probability is termed the neighborhood influence response function $F(m,k)$ \cite{WattsDodds07,LopezPintadoWatts08}; (iii) for any fixed degree $k$, $F(m,k)$ is a nondecreasing function of $m$; and (iv) once active, a node cannot become deactivated. The list of processes which satisfy these constraints includes, but is not necessarily limited to, site and bond percolation \cite{BroadbentHammersley57,StaufferAharony92}, $k$-core decomposition \cite{GoltsevEtAl06,DorogovtsevEtAl06}, and Watts' threshold model \cite{Watts02}. Each process is defined by choosing an appropriate $F(m,k)$, as detailed in \cite{Gleeson08}.

As well as determining the expected cascade size we also provide a cascade condition---that is, a condition specifying the circumstances under which the number of nodes active in the cascade will correspond to a non-vanishing fraction of the total number of nodes in the network $n$ (in the limit $n\rightarrow\infty$). The dependence of such a condition on the prescribed level of clustering has been the topic of much recent discussion \cite{Miller09a,GleesonEtAl10,IkedaEtAl10,Centola10}. The main question under consideration is: ``Does the presence of clustering in $p_{st}$ networks increase or decrease the expected cascade size relative to its value in a nonclustered network with the same degree distribution?'' We provide a general criterion to answer this question.

We restrict our attention throughout to cascades on undirected networks; however, in theory our method should be extendable to directed networks \cite{Gleeson08a}. We also note that while the generating function method of \cite{Newman09} has the added advantage over our approach that it can be used to calculate the entire distribution of cascade sizes, such an approach is not directly generalizable to the wider class of cascade processes considered here.

The remainder of this paper is structured as follows. In Sec.~\ref{sec2}. we describe our generalized approach to cascade dynamics on Newman's clustered random networks. An analytical expression for the mean cascade size and the cascade condition are derived. We define these in terms of an arbitrary response function $F(m,k)$. The particular forms which these results take for various processes are given in Sec.~\ref{sec3} and we discuss in detail the site percolation problem and Watts' threshold model \cite{Watts02}. We investigate the relationship between clustering and the cascade condition in Sec.~\ref{sec4}.

\section{Cascade Propagation}\label{sec2}
Our task here is to show how the theory developed in \cite{GleesonCahalane07,Gleeson08} for cascades on locally tree-like networks can be modified such that it is applicable to the class of clustered random networks introduced in \cite{Newman09}.

Let us begin by recalling some of the properties of that class. First, each network realization is defined by a joint
distribution $p_{st}$ specifying the fraction of nodes connected to $s$ single edges and $t$ triangles. The
conventional degree of each node is, therefore, $k=s+2t$ and the degree distribution is
\begin{equation}
p_{k}=\sum_{s,t=0}^{\infty}p_{st}\dl_{k,s+2t},
\label{eq1}
\end{equation}
where $\dl_{i,j}$ is the Kronecker delta. Second, the clustering coefficient $C$, following the definition given in \cite{NewmanEtAl01}, is
\begin{equation}
C=\frac{3\times(\textrm{number of triangles in network})}{(\textrm{number of connected triples})}
=\frac{3N_{\triangle}}{N_{3}},
\label{eq2}
\end{equation}
where $3N_{\triangle}=n\sum_{st}tp_{st}$ and $N_{3}=n\sum_{k}\binom{k}{2}p_{k}$. Notice that upon substitution
into Eq.~(\ref{eq2}) the factors of $n$ cancel, allowing $C$ to remain nonzero even as $n \rightarrow \infty$.

Now, turning to the theoretical analysis presented in \cite{GleesonCahalane07,Gleeson08}, we see that this was built entirely on the fact that the networks being considered were nonclustered, and so could each be well approximated by a tree in which connections extended strictly from level to level starting from an arbitrary root node. This then allowed the propagation of a cascade to be modeled as a consecutive sequence of activations from a random \emph{child} node on one level to its \emph{parent} node on the next highest level. From a seed fraction of active nodes, the expected size of the ensuing cascade was found by iterating a simple recurrence relation to convergence and then calculating the probability of activation of the root node (see Eqs.~(1)-(3) of \cite{GleesonCahalane07}).

\begin{figure}[h]
\centering \includegraphics[]{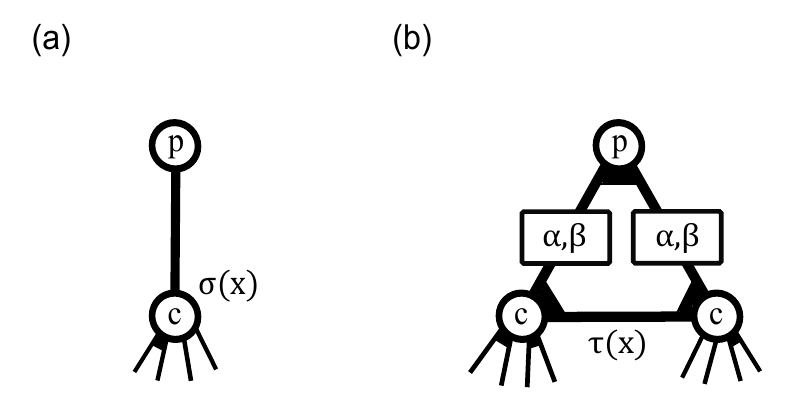}
\caption{Level-by-level cascade propagation in a $p_{st}$ network using the tree approximation. Triangle corners are marked in black.}
\label{fig1}
\end{figure}

If we are to expand this approach to $p_{st}$ networks we must first justify the use of the tree approximation in the presence of nonzero clustering. Observe, however, that in these networks clustering is generated solely through the motif of nonoverlapping triangles. Fitting this specific type of clustering into the tree-based framework is straightforward; a triangle exists whenever an edge connects two nodes on the same level. Therefore, in terms of dynamics, the only difference from the nonclustered networks dealt with in \cite{GleesonCahalane07,Gleeson08} is that now we are faced with two distinct ways in which activations may propagate from one level to the next, see Fig.~\ref{fig1}. They may spread as in Fig.~\ref{fig1}(a) from a child ($c$) to its parent ($p$) across a single edge or as in Fig.~\ref{fig1}(b) from either child at the base of a triangle to the parent at its apex.

\subsection{Expected cascade size}
Following the methodology of \cite{GleesonCahalane07,Gleeson08} then, let us model a generalized cascade as a recursive sequence of activations from child to parent and set up self-consistent equations for the probabilities involved.

Considering first Fig.~\ref{fig1}(a), let $\sg_{1}$ be the probability that the child is active conditional on its parent being inactive, and let $\sg_{0}=1-\sg_{1}$ be the corresponding conditional probability that the child is inactive. For convenience we represent this set of probabilities with the generating function $\sg(x)=\sg_{0}+\sg_{1}x$. Similarly, in Fig.~\ref{fig1}(b), let $\tu_{2}$ be the probability that both children are active, conditional on their parent being inactive, let $\tu_{1}$ be the conditional probability that only one child is active, and let $\tu_{0}=1-\tu_{1}-\tu_{2}$ be the conditional probability that neither child is active. The generating function for these probabilities is $\tu(x)=\tu_{0}+\tu_{1}x+\tu_{2}x^2$.

Of course, the node arrangements represented by Figs.~\ref{fig1}(a)-(b) usually exist in various combinations, and not exclusively of each other. By definition, in any given network realization a randomly chosen node will be directly connected to $s$ nodes via single edges and to $2t$ nodes via triangle edges, with probability $p_{st}$. Therefore, letting $\Pi_{m}^{s,t}$ be the probability that $m$ of these neighboring nodes are active, $\sg(x)$ and $\tu(x)$ are related to that probability by the generating function
\begin{equation}
G(x)=\sum_{m=0}^{s+2t}\Pi_{m}^{s,t}x^{m}=\big[\sg(x)\big]^s\big[\tu(x)\big]^t,
\label{eq3}
\end{equation}
defined for each pairing of $s$ and $t$.

We are now in a position to write down an analytical expression for $\sg_{1}$. In terms of an arbitrary response function $F(m,s+2t)$, written $F_{m}$ for short, we have
\begin{align}
\sg_{1}=\rho_{0}+(1-\rho_{0})\sum_{s,t=0}^{\infty}\frac{sp_{st}}{\la s\ra}\sum_{m=0}^{s+2t-1}\Pi_{m}^{s-1,t}F_{m},
\label{eq4}
\end{align}
where $\rho_{0}$ is the seed fraction and $\la s\ra=\sum_{s,t}sp_{st}$ is the average number of single edges per node. Eq.~(\ref{eq4}) is a self-consistent equation for $\sg_{1}$ since according to Eq.~(\ref{eq3}), $\Pi_{m}^{s,t}$ is itself a function of the coefficients of $\sg(x)$ and $\tu(x)$. We can read Eq.~(\ref{eq4}) as follows: the probability of the child node in a randomly chosen single edge pair being active, conditional on its parent being inactive, is equal to the probability that it was either initially active ($\rho_{0}$), or that ($1-\rho_{0}$) it subsequently became active by copying the behavior of the $m$ out of $s+2t-1$ of its own children that were already active. Note, the term $sp_{st}/\la s\ra$ is the probability of reaching a child with $s$ single edges by traveling along a random single edge from its parent (see \cite{Newman03}).

To obtain similar expressions for $\tu_{1}$ and $\tu_{2}$ we must reflect the fact that in a triangle the state of either child may influence the state of the other. Referring to Fig.~\ref{fig1}(b), the probability that one child is active regardless of the state of the other is
\begin{equation}
\al=\rho_{0}+(1-\rho_{0})\sum_{s,t=0}^{\infty}\frac{tp_{st}}{\la t\ra}\sum_{m=0}^{s+2(t-1)}\Pi_{m}^{s,t-1}F_{m},
\label{eq5}
\end{equation}
the probability that one child is inactive if the other is inactive but will activate if the other is active is
\begin{equation}
\bt=(1-\rho_{0})\sum_{s,t=0}^{\infty}\frac{tp_{st}}{\la t \ra}\sum_{m=0}^{s+2(t-1)}\Pi_{m}^{s,t-1}\big[F_{m+1}-F_{m}\big],
\label{eq6}
\end{equation}
and finally the probability that one child is inactive even if the other is active is $\gm=1-\al-\bt$. In Eqs.~(\ref{eq5})-(\ref{eq6}), we use the fact that following a triangle edge from the parent leads to a child with $t$ triangles with probability $tp_{st}/\la t\ra$. This child then has $s$ single edges and $t-1$ triangles available to connect to its own children, giving its maximum number of active children (for the sum over $m$) as $s+2(t-1)$. Expressed in terms of the probabilities $\al$ and $\bt$, self-consistent expressions for $\tu_{1}$ and $\tu_{2}$ are given by
\begin{equation}
\tu_{1}=2\al\gm,
\label{eq7}
\end{equation}
and
\begin{equation}
\tu_{2}=\al^2+2\al\bt.
\label{eq8}
\end{equation}
The form of Eq.~(\ref{eq7}) arises from the fact that the probability of the parent in a triangle of nodes having one active child is equal to the probability that one child is active regardless of the state of the other ($\al$), while the other is inactive regardless of the state of the other ($\gm$), and there are two different ways in which this may be the case. Reading Eq.~(\ref{eq8}) in the same way, we see that the probability of the parent node in a triangle having two active children is equal to the probability that both children are active regardless of each others' states (${\al}^{2}$) plus the probability that one child is active and the other activates because of this with probability $\bt$, again there are two ways in which the latter may occur.

The propagation of a cascade through a $p_{st}$ network is now almost fully defined. Given a seed fraction $\rho_{0}$, we solve Eqs.~(\ref{eq3})-(\ref{eq8}) to find the steady-state values of the coefficients of the polynomials $\sg(x)$ and $\tu(x)$, and then, using these, we determine the expected cascade size by calculating the probability of activation of the root node. This final probability is given by
\begin{equation}
\rho=\rho_{0}+(1-\rho_{0})\sum_{s,t}^{\infty}p_{st}\sum_{m=0}^{s+2t}\Pi_{m}^{s,t}F_{m}.
\label{eq9}
\end{equation}
Comparing this equation to Eq.~(\ref{eq4}) we see that here the root node, which has $s$ single edges and $t$ triangles with probability $p_{st}$, has no parent and so has $s+2t$ children.

In Sec.~\ref{sec4} we show that the analytical approach derived here is in excellent agreement with the results of numerical simulations on $p_{st}$ networks.

\subsection{Cascade condition}
Having established an analytical expression for the expected cascade size in Eq.~(\ref{eq9}), we now turn to the derivation of a cascade condition. This will determine the circumstances under which the process of propagating activations described by Eqs.~(\ref{eq3})-(\ref{eq8}) can generate a nonvanishing mean cascade size from an infinitesimally small seed fraction $\rho_{0}\rightarrow0$.

We begin by observing that Eqs.~(\ref{eq3})-(\ref{eq8}) can be represented as the steady state of a nonlinear system of the general form $\mathbf{v}^{(n+1)}=\mathbf{H}\big(\mathbf{v}^{(n)}\big)$, where $\mathbf{v}^{(n)}=\big[{\sg_{1}}^{(n)},{\tu_{1}}^{(n)},{\tu_{2}}^{(n)}\big]$. The trivial solution $\mathbf{v}=\mathbf{0}$ corresponds to an equilibrium state where cascades do not occur. We can look for other solutions by applying a small perturbation away from this equilibrium and then considering the trajectories in a linearized version of the system.

Applying this method we first linearize the generating function $G(x)$ of Eq.~(\ref{eq3}) about $\mathbf{v}=\mathbf{0}$ using a small parameter $\ep$ to measure the magnitude of the perturbation. Scaling the coefficients of $\sg(x)$ and $\tu(x)$ as $\mathcal{O}(\ep)$, that is $\sg_{1}\simeq\ep\wt{\sg_{1}}$, $\tu_{1}\simeq\ep\wt{\tu_{1}}$ and $\tu_{2}\simeq\ep\wt{\tu_{2}}$, we expand $G(x)$ as
\begin{equation}
G(x)\simeq1-\ep\big[s\wt{\sg_{1}}+t(\wt{\tu_{1}}+\wt{\tu_{2}})-(s\wt{\sg_{1}}+t\wt{\tu_{1}})x-t\wt{\tu_{2}}x^2\big],
\label{eq10}
\end{equation}
up to terms of $\mathcal{O}(\ep^2)$.

Our next step will be to substitute the coefficients of $G(x)$ from Eq.~(\ref{eq10}) into Eqs.~(\ref{eq4})-(\ref{eq8}). Before doing this, however, we further simplify our analysis by assuming $F_{0}=0$.
This implies that a node will never activate if none of its neighbors are active, and this is true, or a good approximation, in many cases of interest.
With $F_{0}=0$ then, said substitution gives us a linear system which may be represented in the matrix form $\wt{\mathbf{v}}^{(n+1)}=\mathbf{A}\cdot \wt{\mathbf{v}}^{(n)}$, where
\begin{eqnarray}
\wt{\bf{v}}^{(n)}=\left[ \begin{array}{c} \wt{\sg_{1}}^{(n)} \\ \wt{\tu_{2}}^{(n)} \end{array}\right],
\bf{A}=\begin{bmatrix} A_{11} & A_{12} \\ A_{21} & A_{22} \end{bmatrix},
\label{eq11}
\end{eqnarray}
and the elements of $\mathbf{A}$ are
\begin{eqnarray}
\nn &A_{11}=\frac{\la (s^2-s)F_{1} \ra}{\la s \ra}, A_{12}=\frac{\la stF_{2}\ra}{\la s\ra}+\frac{\la stF_{1}\ra}{\la s\ra}\frac{\la t\ra-\la tF_{1}\ra}{\la tF_{1}\ra};\\
\nn &A_{21}=\frac{2\la stF_{1}\ra\la tF_{1}\ra}{\la t\ra^2},\\
&A_{22}=\frac{2\la (t^2-t)F_{1} \ra}{\la t \ra}+\frac{2\la (t^2-t)(F_{2}-F_{1})\ra\la tF_{1}\ra}{\la t\ra^2}.
\label{eq12}
\end{eqnarray}
Note, the application of Eq.~(\ref{eq10}) has allowed us to express $\wt{\tu_{1}}^{(n)}$ in terms of $\wt{\tu_{2}}^{(n)}$ as $\wt{\tu_{1}}^{(n)}=(\la t\ra-\la tF_{1}\ra)\wt{\tu_{2}}^{(n)}/\la tF_{1}\ra$, hence the reduction to the $2\times2$ system of linear equations represented by Eqs.~(\ref{eq11})-(\ref{eq12}).

In order for this system to produce trajectories which will diverge from $\mathbf{v}=\mathbf{0}$, in other words in order to produce cascades, we require that the larger eigenvalue of $\mathbf{A}$ (both eigenvalues are real) be greater than one, $\lambda_{+}>1$ \footnote{In general, the trivial equilibrium $\mathbf{v}=\mathbf{0}$ is unstable when the matrix $\mathbf{A}$ has at least one eigenvalue $\lambda$ such that $\left|\lambda\right|>1$. Our simplified requirement follows from the fact that $\mathbf{A}$ consists of real positive elements and thus according to the Perron-Frobenius theorem at least one of the eigenvalues of $\mathbf{A}$ is real and positive, and is greater than the other in absolute value.}. This condition is satisfied if
\begin{eqnarray}
\nn &\la t\ra\Big[2\la st F_{1} \ra^2-\big(\la(s^2-s)F_{1}\ra-\la s\ra\big)\big(2\la(t^2-t)F_{1}\ra-\la t\ra\big)\Big]\\
\nn &-2\la t F_{1}\ra\Big[\big(\la(s^2-s)F_{1}\ra-\la s\ra\big)\la(t^2-t)(F_{2}-F_{1})\ra\\
&-\la st F_{1}\ra\la st(F_{2}-F_{1})\ra\Big]>0.
\label{eq13}
\end{eqnarray}
Conversely, if the left hand side of Eq.~(\ref{eq13}) is negative then $\lambda_{+}<1$, and the trivial equilibrium is stable, so cascades do not occur. The boundary between these two regimes, one where cascades are observed and the other where they are not, is located precisely at the point where $\lambda_{+}=1$, or equivalently where the expression on the left hand side of Eq.~(\ref{eq13}) is equal to zero.

\section{Response Functions}\label{sec3}
In this section we will show how the generalized theory of Sec.~\ref{sec2} may be used to model a range of processes on $p_{st}$ networks. As stated in the introduction each specific process will be defined by choosing an appropriate response function, and Eqs.~(\ref{eq3})-(\ref{eq9}) will then give the expected cascade size. We consider the examples of site percolation and Watts' threshold model \cite{Watts02} in detail.

\subsection{Site and bond percolation}
The resilience of random networks in the face of indiscriminate breakdowns or coordinated attacks is a key concern across multiple disciplines from epidemiology to telecommunications. Modeling these types of events as percolating processes has proved to be very fruitful, allowing theorists to uncover formulas for, amongst other things, the size distribution of connected components \cite{CallawayEtAl00,Newman07} and epidemic thresholds \cite{Grassberger83}. The two most basic models studied are uniform site percolation and uniform bond percolation. Here we show that both models may be considered as special cases of our generalized approach, corresponding to suitable choices for the response function $F(m,s+2t)$.

Following the approach of \cite{GleesonMelnik09}, we frame our description in the language of successive activations already introduced. We define a node as active if it is part of the giant connected component (GCC) of the network, and our choice of response function, Eq.~(\ref{eq14}) or Eq.~(\ref{eq18}) below, determines the type of percolation under  consideration, either site percolation or bond percolation respectively. When this activation process reaches steady-state, all nodes which are labeled as active have at least one active neighbor to which they are connected. Thus the fraction $\rho$ of active nodes equals the size of the connected component, expressed as a fraction of the network size $n$. In the $n\to\infty$ limit, only the giant connected component size scales with $n$, and so $\rho$ gives the fractional size of the GCC. This can be seen also from the fact that in the limit of zero clustering, our equations reduce to the standard percolation equations for GCC size in configuration model networks, as given in \cite{CallawayEtAl00}. This method does not permit the calculation of finite-size connected components (see \cite{Gleeson08,Gleeson09,GleesonMelnik09}).

In uniform site percolation, each node is occupied with independent probability $\mu$ and an occupied node can become active in the cascade, i.e. form part of the giant connected component (GCC), if it has one or more active neighbors (who are already in the GCC). Unoccupied nodes can never become active. The response function for site percolation is therefore \cite{Gleeson08},
\begin{align}
&F(m,s+2t)=\begin{cases}
0 &\text{if} \quad m=0,\\
\hfill \mu &\text{if} \quad m>0.\\
            \end{cases}
\label{eq14}
\end{align}

Using Eq.~(\ref{eq14}) in the $\rho_{0}\rightarrow0$ limit of Eqs.~(\ref{eq4})-(\ref{eq9}), and noting that with this choice of response function
\begin{equation}
\sum_{m=0}^{s+2t}\Pi_{m}^{s,t}F(m,s+2t)=\mu\Big[1-{\sg_{0}}^{s}{\tu_{0}}^{t}\Big],
\label{eq15}
\end{equation}
the expected size of the GCC (as $n\rightarrow\infty$) is given by Eq.~(\ref{eq9}), and reduces to the simple form
\begin{equation}
\rho=\mu-\mu\sum_{s,t=0}^{\infty}p_{st}{\sigma_{0}}^{s}{\tau_{0}}^{t}.
\label{eq16}
\end{equation}
Substituting Eq.~(\ref{eq14}) into our cascade condition Eq.~(\ref{eq13}) we derive the following equation for the critical site percolation occupation probability
\begin{equation}
\big(\mu\la s^2-s\ra-\la s\ra\big)\big(2\mu\la t^2-t\ra-\la t\ra\big)
-2{\mu}^2\la st \ra^2=0,
\label{eq17}
\end{equation}
which, with $\mu=1$, is in agreement with Eq.~(22) of \cite{Newman09}.

In uniform bond percolation each edge is occupied with independent probability $\nu$ and a node can become active only if it is linked to another active node by an occupied edge. Thus, a node with $m$ active children has probability $1-(1-\nu)^m$ of becoming active itself. The appropriate choice of response function in this case is therefore \cite{Gleeson08},
\begin{align}
&F(m,s+2t)=\begin{cases}
0 &\text{if} \quad m=0,\\
\hfill 1-(1-\nu)^{m} &\text{if} \quad m>0.\\
            \end{cases}
\label{eq18}
\end{align}

The approach outlined here is also applicable to two other closely related problems: susceptible-infected-recovered (SIR) disease transmission \cite{Grassberger83,Newman02} and $k$-core decomposition \cite{GoltsevEtAl06,DorogovtsevEtAl06}. In fact, it was shown in \cite{Newman02} that in the steady state the infected fraction in SIR may be mapped directly to the bond percolation problem. The latter was discussed in detail in \cite{Gleeson08} and the relevant response function for standard configuration model networks was provided (see Eq.~(10) of \cite{Gleeson08}). With the introduction of triangles we simply update that response function $F(m,k)$ by setting $k=s+2t$ and continue as above.

Regarding epidemiological studies, the question of how clustering in networks of human interactions may influence the size and persistence of outbreaks of infectious diseases has been the topic of much recent discussion \cite{Trapman07,Eames08,BrittonEtA08,Miller09b,BallEtAl10,HebertDufresneEtAl10}. In fact, much of the impetus for considering more complex topological motifs in studies involving networked structures in general has come from this source \cite{Trapman07,GleesonMelnik09}. We will see in Sec.~\ref{sec4} how the results obtained by us for site and bond percolation echo (albeit indirectly) a number of recent results from this literature concerning the effects of clustering.

\subsection{Watts' model}
In \cite{Watts02} Watts introduced a model of threshold dynamics on networks as a simple but plausible mechanism for how phenomena such as fads or rumors propagate in society. The most basic formulation of this model is as follows. In an undirected network of arbitrary degree distribution $p_{k}$, assign to each node a random (frozen) threshold $r$ drawn from a specified distribution. Then, starting from a small seed fraction of active nodes, $\rho_{0}$, synchronously update the state of each node based on the following decision rule: a node will become active if the fraction of its neighbors which are already active exceeds $r$, otherwise it will remain inactive. (We also stipulate that once active a node can not deactivate.) Repeating this updating process until a steady-state is reached, we call the final fraction of active nodes the cascade size.

In \cite{Gleeson08} Gleeson defined the response function for Watts' model in the context of a generalized approach to cascades on $p_{k}$ networks (see Eq.~(2) of \cite{Gleeson08}). We can extend this definition to $p_{st}$ networks simply by setting $k=s+2t$. From Eq.~(2) of \cite{Gleeson08} this gives us
\begin{equation}
F(m,s+2t)=C_{r}\Bigg(\frac{m}{s+2t}\Bigg),
\label{eq19}
\end{equation}
where $m$ is the number of active neighbors and $C_{r}$ denotes the cumulative distribution function (cdf) of the thresholds. If, for example, we require a Gaussian threshold distribution with mean $R$ and standard deviation $\sg$, then Eq.~(\ref{eq19}) becomes
\begin{equation}
F(m,s+2t)=\frac{1}{2}\Bigg[1+\textrm{erf}\Bigg(\frac{m/(s+2t)-R}{\sg\sqrt{2}}\Bigg)\Bigg],
\label{eq20}
\end{equation}
where erf$(x)$ is the error function. Note, $F(0,s+2t)>0$ here, meaning some nodes have negative thresholds, and so will activate even if none of their neighbors are active. It is possible, therefore, for such nodes to instigate a cascade even when $\rho_{0}=0$.

In a similar manner to before we obtain the mean cascade size and the cascade condition by substituting Eq.~(\ref{eq19}) into the relevant equations from Sec.~\ref{sec2}.

\section{Effects of Clustering on Cascades}\label{sec4}
We now turn to the investigation of how clustering can affect cascade dynamics on $p_{st}$ networks. This requires first that we make an appropriate choice for the form of the joint distribution $p_{st}$. Considering the question stated in the introduction: ``Does the presence of clustering in $p_{st}$ networks increase or decrease the expected cascade size relative to its value in a nonclustered network with the same degree distribution?'', we set
\begin{equation}
p_{st}=p_{k}\dl_{k,s+2t}\big[(1-f)\dl_{t,0}+f\dl_{t,\lfloor (s+2t)/2\rfloor}\big],
\label{eq21}
\end{equation}
where $f\in[0,1]$, and $\lfloor\cdot\rfloor$ is the floor function.

Applying this definition, we construct $p_{st}$ from a given degree distribution $p_{k}$ such that a fraction $f$ of the nodes in our network are attached to the maximum possible number of triangles $t=\lfloor(s+2t)/2\rfloor$ while the remaining $(1-f)$ are attached to single edges only ($t=0$). Upon substitution of Eq.~(\ref{eq21}) into Eq.~(\ref{eq2}) we find that the clustering coefficient $C$ can be expressed as
\begin{equation}
C=f\frac{\sum_{k}k(p_{2k}+p_{2k+1})}{\sum_{k}\binom{k}{2}p_{k}}.
\label{eq22}
\end{equation}
This linear relationship between $C$ and $f$ allows us to vary $C$ continuously from its minimum value at $f=0$ to its maximum possible value obtained at $f=1$, while preserving $p_{k}$ throughout. We cannot guarantee, however, that degree-degree correlations will be preserved \cite{GleesonEtAl10}.

\begin{figure}[t]
\centering \includegraphics[width=8.8cm]{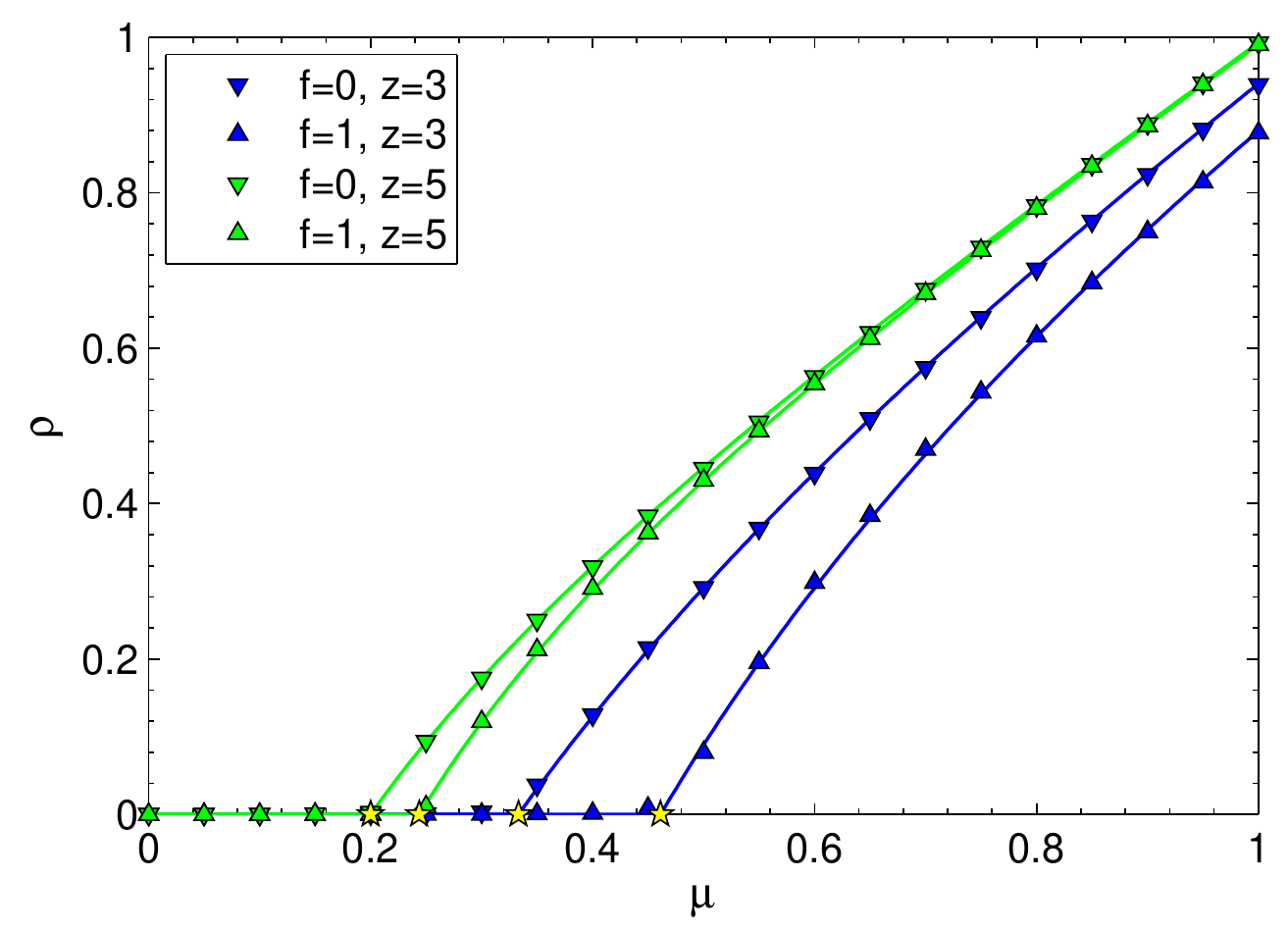}
\caption{(Color online) Size of giant connected component $\rho$ as a function of site occupation probability $\mu$ on $p_{st}$ networks of $10^{5}$ nodes with Poisson degree distribution $p_{k}$ for two different
values of the mean degree, $z=3$ and $z=5$. Numerical simulations averaged over $100$ realizations (symbols) versus theory of Sec.~\ref{sec2} (solid lines). In both cases we consider minimum clustering $f=0$ and
maximum clustering $f=1$. In each of the four parameter settings we calculate the critical site occupation probability from Eq.~(\ref{eq17}) and mark its position on the $\mu$ axis with a (yellow) pentagram.}
\label{fig2}
\end{figure}

In Fig.~\ref{fig2} we have used Eq.~(\ref{eq21}) to verify our theory in the case of site percolation on $p_{st}$ networks with Poisson degree distribution $p_{k}=z^{k}e^{-z}/k!$. We plot our result for the GCC size from Eq.~(\ref{eq16}) against numerical simulations for two different values of the mean degree $z=\sum_{k}kp_{k}$. In both cases we consider minimum clustering ($f=0$) and maximum clustering ($f=1$). Threshold values defined by Eq.~(\ref{eq17}) are also plotted (see caption for details).

Observing the relative positions of the percolation thresholds in Fig.~\ref{fig2} (pentagrams) we note that they lend support in favor of (or, at least, do not contradict) the argument that adding triangles decreases the cascade size. We showed in \cite{GleesonEtAl10} that this is unambiguously the case in the bond percolation problem on $z$-regular $p_{st}$ networks, i.e. those with $p_{k}=\dl_{k,z}$ (all nodes have $z$ neighbors). However, since adding triangles to a $z$-regular network cannot affect its correlation structure, this meant that any effects which may have been introduced by allowing correlations to vary were automatically negated. Furthermore, it was explicitly demonstrated in \cite{GleesonEtAl10}, and also \cite{Miller09a}, that such effects may significantly complicate matters. In Fig.~\ref{fig2}, on the other hand, degree-degree correlations are not preserved. Therefore while this figure does validate the theoretical approach of the preceding sections it does not permit us to draw definitive conclusions as regards the question of the change in the expected cascade size due to clustering alone.

In order to do that we will follow the approach of \cite{GleesonEtAl10} (see also \cite{Centola10}) and focus our investigation on $p_{st}$ networks with $z$-regular $p_{k}$. In particular, we consider the following joint distribution

\begin{equation}
p_{st}=\dl_{z,s+2t}\big[(1-g)\dl_{t,0}+g\dl_{t,1}\big],
\label{eq23}
\end{equation}
where $z>2$. This choice shares some similarities with Eq.~(\ref{eq21}); however, here we are adding only one triangle to each of a fraction $g$ of the nodes in a $z$-regular network. Substituting Eq.~(\ref{eq23}) into Eq.~(\ref{eq13}) we have, as the condition for cascades to occur (corresponding to $\lambda_{+}>1$, see Sec.~\ref{sec2}(B)),
\begin{equation}
F_{1}(z^2-z)-z+gS_{c}>0,
\label{eq24}
\end{equation}
where
\begin{align}
\nn S_{c}=2&+F_{1}(6-4z)+2{F_{1}}^2(z-2)^2\\
&+2{F_{1}}^2F_{2}(z-2)^2-2{F_{1}}^3(z-2)^2,
\label{eq25}
\end{align}
denotes the sum of the terms which introduce clustering into the network. This expression gives us an insight
into how adding triangles alters the cascade size. Given a specific $z$ we can determine the qualitative effect of clustering in the
following way. First, set the expression on the left hand side of Eq.~(\ref{eq24}) equal to zero and solve for $F_{1}$ at $g=0$. This
determines the value of $F_{1}$ at the transition to the cascade regime in the nonclustered network; the well-known result of
Watts \cite{Watts02}, $F_{1}=1/(z-1)$. Next, substitute that $F_{1}$ into $S_{c}$ and observe its sign. If it is negative
we conclude that introducing triangles will decrease the expected cascade size. If, on the other hand, $S_{c}$
is positive, adding triangles will increase the cascade size.

The justification for these last two statements follows simply from the fact that if $S_{c}$ constitutes a negative contribution to the
expression on the left hand side of Eq.~(\ref{eq24}) then increasing $g$, given that $F_{1}=1/(z-1)$, will break the inequality in
Eq.~(\ref{eq24}) and take us into the regime where cascades do not occur. Alternatively, if $S_{c}$ is shown to be positive then
increasing the parameter $g$ will ensure the inequality holds and cascades do occur at these parameter values.

\begin{figure}[t]
\centering
\includegraphics[width=8.8cm]{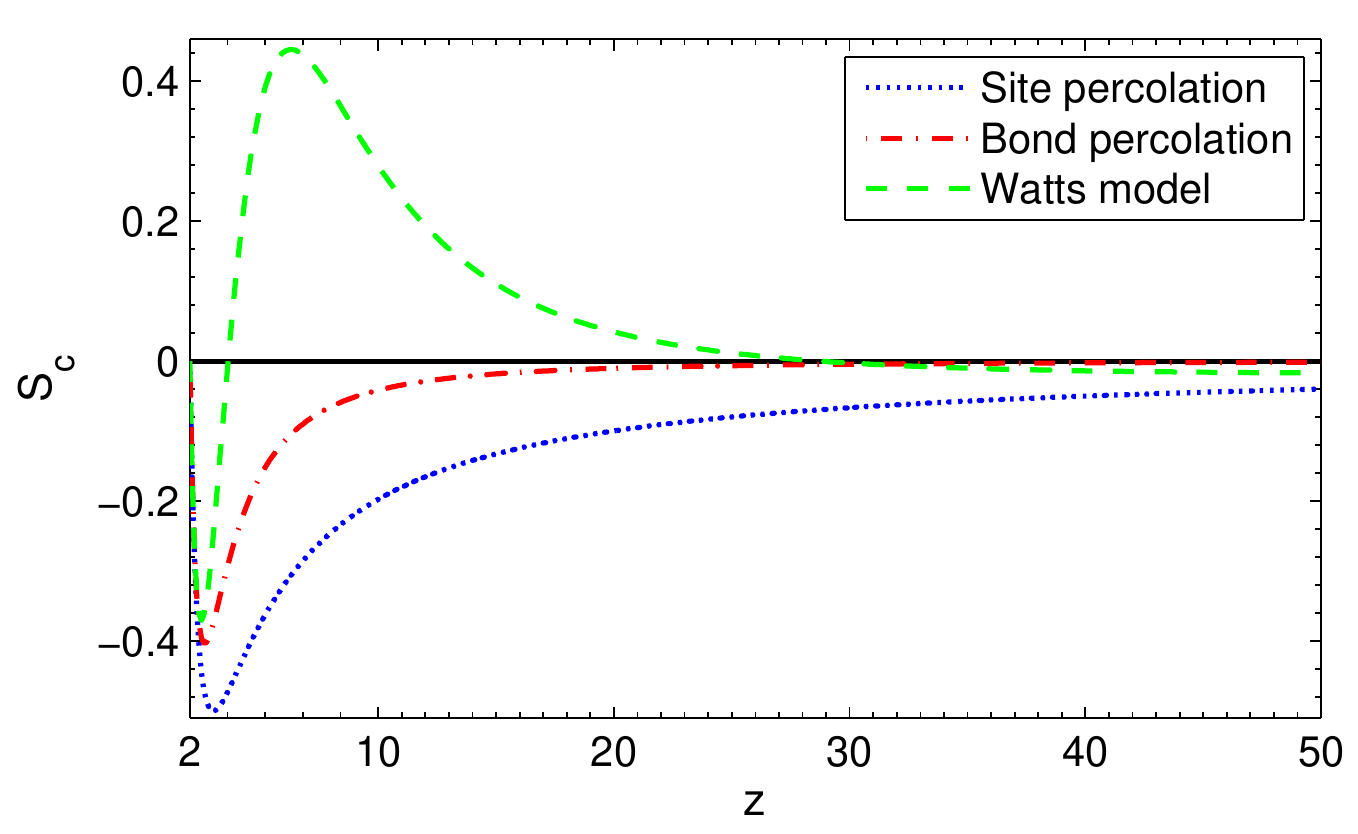}
\caption{(Color online) Sum of the clustering terms from Eq.~(\ref{eq24}), $S_{c}$, versus mean degree $z$ on $p_{st}$ networks with $z$-regular degree distribution. Results from site percolation, bond percolation, and Watts' model are shown. As in Sec.~\ref{sec3}, each process is defined by choosing an appropriate response function. For Watts' model the threshold distribution is Gaussian with standard deviation $\sg=0.1$ and mean $R$, such that $F_{1}=1/(z-1)$. Note, only integer $z$ values are realizable as $z$-regular networks.}
\label{fig3}
\end{figure}

In Fig.~\ref{fig3} we have plotted $S_{c}$ against $z$ for three of the processes described in Sec.~\ref{sec3}: site percolation, bond percolation and Watts' model. In this last case we have chosen the following parameters; seed fraction $\rho_{0}=0$, and a Gaussian threshold distribution with mean $R$ fitted to $F_{1}=1/(z-1)$, and standard deviation $\sigma=0.1$.

This plot indicates that adding triangles will decrease the expected cascade size in both site percolation and bond percolation. In other words, the occupation probability needed for a giant connected component to exist (the percolation threshold) is increased in the presence of clustering. As mentioned above, we have already demonstrated in \cite{GleesonEtAl10} that this is the case for the latter of these two processes; to our knowledge this is the first statement of the corresponding result for site percolation. While these results are not directly applicable to models of the spread of disease, in light of the established connection between SIR epidemics and bond percolation we suggest that they may, nonetheless, be of some interest to researchers in that field. This statement is vindicated by the fact that analogous results have recently been established in a number of epidemiological studies which have shown that clustering can adversely affect the propagation of a disease \cite{Eames08,Miller09b,BallEtAl10,HebertDufresneEtAl10}.

Also of interest is the behavior of $S_{c}$ for Watts' model. As $z$ increases in Fig.~\ref{fig3}, we see $S_{c}$ vary from negative values for $z\leq3$, through a regime of positivity, and back again to negative values for $z\geq29$. This tells us that for $z\leq3$ the presence of clustering will decrease the left hand side of Eq.~(\ref{eq24}) below zero, thereby \emph{decreasing} the expected cascade size; for $3<z<29$ clustering will \emph{increase} the expected cascade size; and finally for $z\geq29$ clustering will once more tend to \emph{decrease} the expected cascade size. We note that qualitatively similar results are seen for different values of $\sg$, the standard deviation of the thresholds.

By way of validation, in Fig.~\ref{fig4} we plot the cascade size $\rho$ against the mean of the threshold distribution $R$ for Watts' model with joint distribution defined by Eq.~(\ref{eq21}), and otherwise the same parameter settings as in Fig.~\ref{fig3} (see caption for details). We inferred from Fig.~\ref{fig3} that at $z=3$ cascades become smaller as clustering is increased. This is what we observe in Fig.~\ref{fig4}(a). Contrastingly, at $z=5$ cascades should become larger as clustering increases. This is verified by Fig.~\ref{fig4}(b).
\begin{figure}[t]
\centering \includegraphics[width=8.8cm]{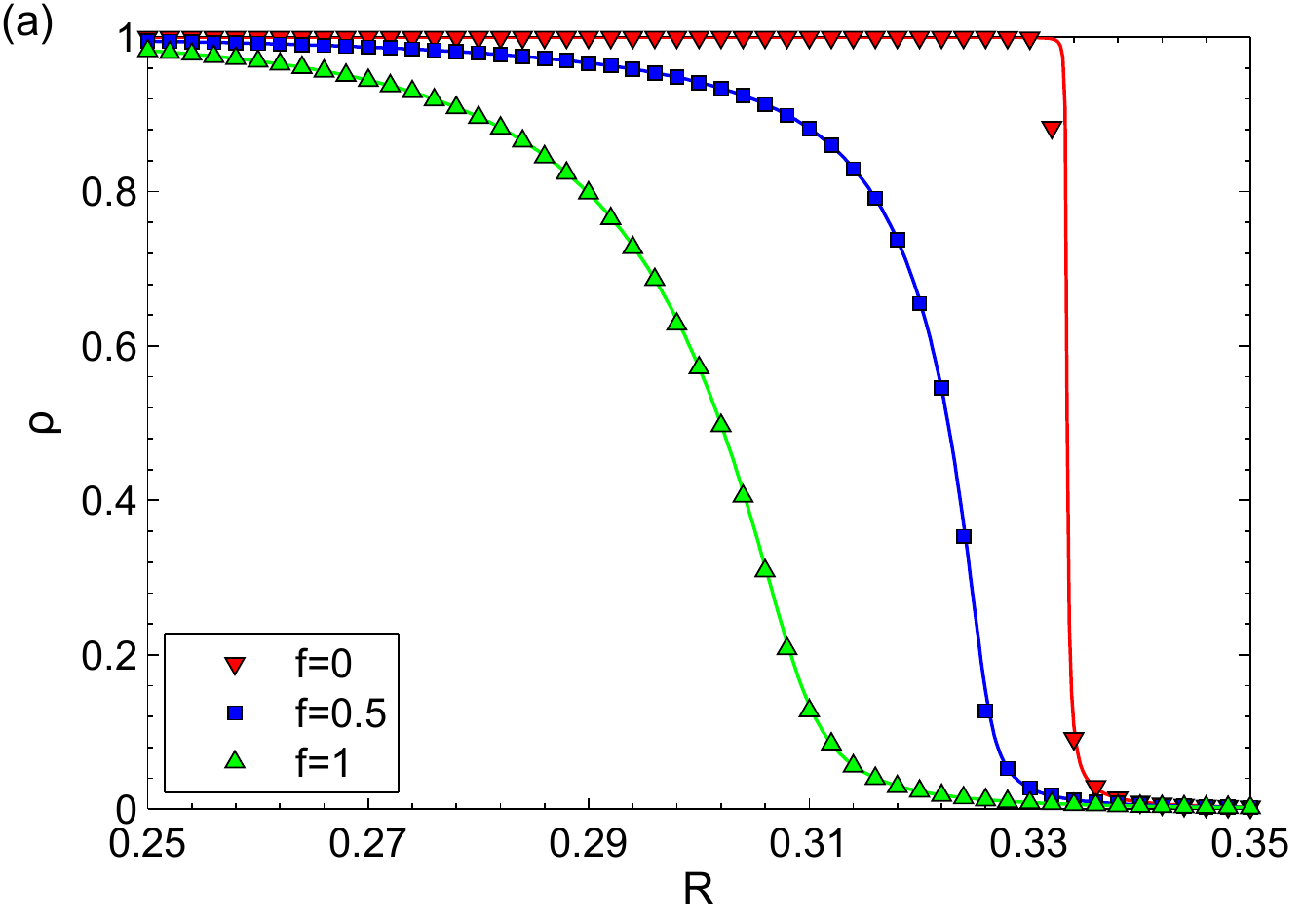}
\end{figure}
\begin{figure}[t]
\centering \includegraphics[width=8.8cm]{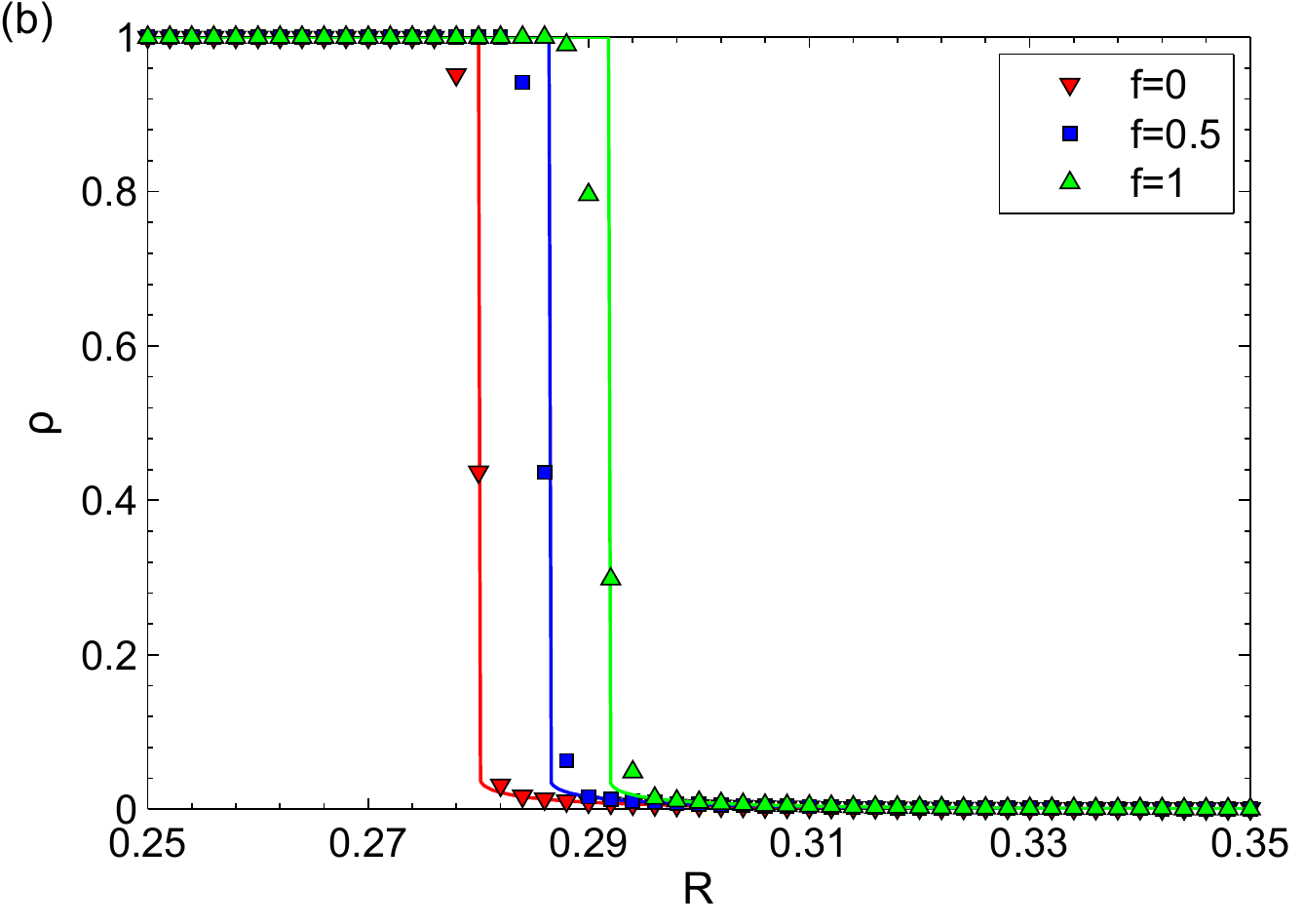}
\caption{(Color online) Expected size of cascade outbreak $\rho$ versus mean $R$ of a Gaussian threshold distribution with $\sg=0.1$ for Watts' model on graphs of $10^{5}$ nodes with $z$-regular degree distribution $p_{k}=\dl_{k,z}$ and joint distribution $p_{st}$ defined by Eq.~(\ref{eq21}). Numerical simulations averaged over $100$ realizations (symbols) and theory of Sec.~\ref{sec2} (solid lines). (a) $z=3$: here increasing the level of clustering decreases the expected cascade size at any given $R$ value; (b) $z=5$: increasing the level of clustering increases the expected cascade size.}
\label{fig4}
\end{figure}

This dependence of the cascade size on the sign of the sum of the clustering terms in Eq.~(\ref{eq24}), $S_{c}$, may be expressed succinctly as a condition on the response function $F_{2}$, the probability of activation in the presence of two active neighbors. Specifically, if the value of $F_{2}$ at the transition point for cascades in nonclustered $z$-regular networks (i.e., $F_{2}$ evaluated at the parameters for which $F_{1}=1/(z-1)$) satisfies the condition
\begin{equation}
F_{2}\Bigg|_{F_{1}=\frac{1}{z-1}}>\frac{2z-3}{(z-2)(z-1)},
\label{eq26}
\end{equation}
then adding triangles will increase the expected size of cascades. Alternatively, if $F_{2}$ does not satisfy this inequality, clustering will decrease the expected size of cascades. One may derive this condition by substituting the zero-clustering cascade condition $F_{1}=1/(z-1)$ into Eq.~(\ref{eq24}) and then solving for $F_{2}$. Note that by substituting the respective response functions for site and bond percolation, Eq.~(\ref{eq14}) and Eq.~(\ref{eq18}), into Eq.~(\ref{eq26}) one may confirm that for $z>2$ this inequality is not satisfied, and thus that clustering decreases cascade sizes for both of these processes (increases the percolation threshold).
Finally, note that Eq.~(\ref{eq26}) can also be arrived at by a simple counting argument which compares the spread of activations in a clustered random network to that in a nonclustered random network. We leave this discussion to the Appendix.

\section{Conclusions}\label{sec5}
We have shown how the analytical approach to cascade dynamics on nonclustered configuration model networks first put forth by Gleeson and Cahalane in \cite{GleesonCahalane07} may be extended to the class of random networks with nonzero clustering described by Newman in \cite{Newman09}.

By adapting the  approach of \cite{Gleeson08} we have provided a general analytical expression for the expected size of a cascade outbreak and a cascade condition, in these more realistic network topologies. By the use of the response function mechanism both of these results may be applied to a range of processes including, but not necessarily limited to, site and bond percolation, $k$-core decomposition, SIR (susceptible-infected-recovered) disease transmission, and Watts' threshold model (see also \cite{BaxterEtAl10}).

In addition to this, we have also considered the question of how the presence of clustering qualitatively affects the cascade condition. This question is further complicated by the fact that for heterogeneous degree distributions, adding triangles will alter the correlation structure of the network \cite{Miller09a,GleesonEtAl10}. We have therefore focused our investigation on clustered networks with $z$-regular degree distributions in which degree-correlation effects are absent. This enabled us to discover a condition on the response function of the process (see Eq.~(\ref{eq26})) which determines the change in the expected size of the cascade due to clustering alone.

For site and bond percolation we found that clustering will unambiguously decrease the cascade size: a result which bears analogy to recent results from the epidemiological literature concerning the effects of clustering on disease outbreaks \cite{Eames08,Miller09b,BallEtAl10,HebertDufresneEtAl10}. For Watts' model, however, matters are not so clear-cut. For certain values of $z$ clustering may increase the mean cascade size, while for others it will decrease it. The example of this behavior provided in Fig.~\ref{fig3}  corresponds to just one setting of parameters for Watts' model, namely a Gaussian threshold distribution with standard deviation $\sg=0.1$ and no seed nodes. We note, however, that further simulations, the results of which are not provided, have proved these observations to be robust against changes in $\sg$. We believe, therefore, that these observations have significant implications for studies of the spread of behavior in social networks, such as for example \cite{Centola10}.

Lastly, we must emphasize that the motif of nonoverlapping triangles in the model investigated here corresponds to just one of the many different ways in which nodes may cluster together in a network. An alternative model based on the idea of embedding \emph{cliques} of nodes within a configuration type network was developed by Gleeson \cite{Gleeson09}, while Karrer and Newman have recently proposed an approach which allows for a much broader range of clustering motifs than just triangles \cite{KarrerNewman10}. The investigation of some of the questions discussed by us here in the context of these models is of significant interest.

\begin{acknowledgements}
This work was funded by Science Foundation Ireland under
programmes 06/IN.1/I366 and MACSI 06/MI/005.
\end{acknowledgements}

\appendix*
\section{Counting Argument for Condition on $F_{2}$}\label{appA}
Here we give an intuitive argument for the effect of clustering on cascades in $z$-regular $p_{st}$ networks. This stands as an alternative derivation of the condition on $F_{2}$ in the main body of this paper, see Eq.~(\ref{eq26}).

\begin{figure}[t]
\centering \includegraphics[scale=0.9]{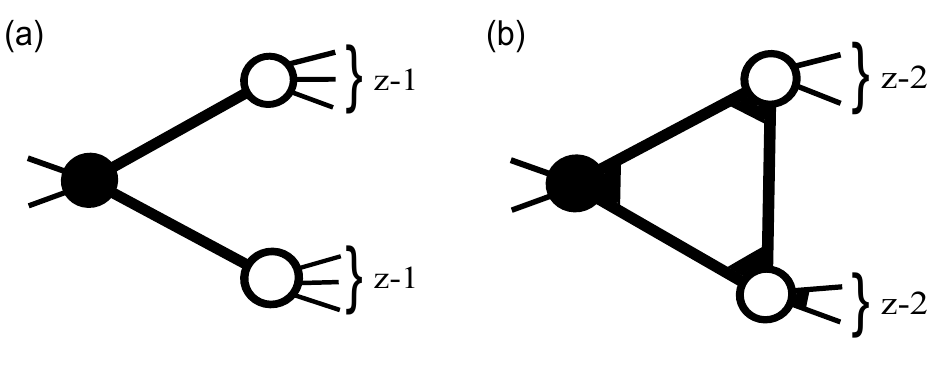}
\caption{Spread of activation from a single node (colored black) in (a) a nonclustered network, and (b) a $p_{st}$ network with nonzero clustering. Note, the clustered $p_{st}$ network may also contain
single edges which are not part of any triangle; however, such edges are also present in the nonclustered network and we are interested only in the differences introduced by adding triangles.}
\label{fig5}
\end{figure}

We compare the spread of activation from a single node (colored black in Fig.~\ref{fig5}(a) and (b)) to two of its neighbors, and then further into the network. In configuration (a) the three nodes considered are not part of a triangle, and up to $2(z-1)$ second neighbors may potentially be activated in this way. In configuration (b), the three nodes form a triangle, and only $2(z-2)$ second neighbors are available for activation. We proceed to calculate the expected number of edges which may activate second neighbors in each configuration, and derive a condition under which clustering (configuration (b)) gives a larger number of expected activations than the corresponding nonclustered case (configuration (a)). First we consider configuration (a). Each of the two white nodes will be activated by the black node with probability $F_{1}$. If activated, a white node may in turn activate up to $z-1$ of its other neighbors. So we count the expected number of \emph{active edges} (edges which are connected to an active node) on the right hand side of Fig.~\ref{fig5}(a) as $2F_{1}(z-1)$.

In configuration (b), the two neighbors of the active node are also connected to each other, leaving each with $z-2$ edges to other neighbors. These edges may become active edges in one of three ways.
\begin{enumerate}[(i)]
  \item Both white nodes are activated directly by their single active neighbor; this happens with probability ${F_{1}}^2$, and gives $2(z-2)$  active edges on the right hand side of Fig.~\ref{fig5}(b).
  \item One white node is activated directly by the active neighbor; the other white node then becomes active because it now has two active neighbors. This happens with probability
  $2F_{1}(F_{2}-F_{1})$, and gives $2(z-2)$ active edges.                                                                                                                                                                        \item One white node is activated directly by the active neighbor; the other white node does not activate even though it has two active neighbors. This happens with probability
  $2F_{1}(1-F_{2})$, and gives $z-2$ active edges.
\end{enumerate}

The expected number of active edges on the right hand side of Fig.~\ref{fig5}(b) is therefore
\begin{align}
\nn 2{F_{1}}^{2}(z-2)&+4F_{1}(F_{2}-F_{1})(z-2)+2F_{1}(1-F_{2})(z-2)\\ &=2F_{1}(z-2)(F_{2}-F_{1}+1).
\label{A1}
\end{align}
This is larger than the value $2F_{1}(z-1)$ found for configuration (a) if
\begin{equation}
F_{2}-F_{1}>\frac{1}{z-2}.
\label{A2}
\end{equation}
To examine the effect upon the cascade threshold, we substitute the cascade condition $F_{1}=1/(z-1)$ for the threshold in a nonclustered $z$-regular network \cite{Watts02} into Eq.~(\ref{A2}) to obtain the condition given in Eq.~(\ref{eq26}). If this condition is satisfied, cascade propagation is more likely on the clustered $z$-regular network than on the nonclustered version.

\bibliography{CCRN_v2}
\end{document}